\DocumentMetadata{}
\documentclass[sigconf]{acmart}

\acmConference[CAIN ’26]
  {5th International Conference on AI Engineering -- Software Engineering for AI}
  {April 12--13, 2026}
  {Rio de Janeiro, Brazil}

\makeatletter
\def\mdseries@tt{m}
\makeatother

\pdfoutput=1

\usepackage{amsmath}
\usepackage{amsfonts}
\usepackage{graphicx}
\usepackage{natbib}
\usepackage{booktabs}
\usepackage{array}
\usepackage{url}
\usepackage{listings}
\usepackage{xcolor}
\usepackage{caption}
\usepackage{subcaption}
\usepackage{float}
\usepackage{geometry}
\usepackage{fancyhdr}
\usepackage{algorithm}
\usepackage{algorithmic}
\usepackage{multirow}
\usepackage{mdframed}
\usepackage{xcolor}
\usepackage{xstring}

\usepackage{tikz}
\definecolor{GoodGreen}{RGB}{0,139,69}
\definecolor{BadRed}{RGB}{204,0,0}

\newcommand{\impact}[2]{%
  \IfStrEq{#2}{red}{%
    \textcolor{BadRed}{\textbf{#1}}%
  }{%
    \IfStrEq{#2}{green}{%
      \textcolor{GoodGreen}{\textbf{#1}}%
    }{%
      \textbf{#1}%
    }%
  }%
}

\begin{document}
\title{Saving SWE-Bench: A Benchmark Mutation Approach for Realistic Agent Evaluation}

\author{Spandan Garg}
\authornote{Corresponding author.}
\email{spgarg@microsoft.com}
\affiliation{%
  \institution{Microsoft}
  \country{USA}
}
\author{Benjamin Steenhoek}
\email{bensteenhoek@microsoft.com}
\affiliation{%
  \institution{Microsoft}
  \country{USA}
}
\author{Yufan Huang}
\email{yufanhuang@microsoft.com}
\affiliation{%
  \institution{Microsoft}
  \country{USA}
}

\date{}

\begin{abstract}
Current benchmarks for evaluating software engineering agents, such as SWE-Bench Verified, are predominantly derived from GitHub issues and fail to accurately reflect how developers interact with chat-based coding assistants in integrated development environments (IDEs). We posit that this mismatch leads to a systematic overestimation of agent's capabilities in real-world scenarios, especially bug fixing. We introduce a novel benchmarking framework that transforms existing formal benchmarks into realistic user queries through systematic analysis of developer interaction patterns with chat-based agents. Our methodology is flexible and can be easily extended to existing benchmarks. In this paper, we apply our testing framework to SWE-Bench Verified, the TypeScript subset of Multi-SWE-Bench and a private benchmark, SWE-Bench C\# and transform formal GitHub issue descriptions into realistic user-style queries based on telemetry analysis of a popular chat-based agent interactions. Our findings reveal that existing benchmarks significantly overestimate agent capabilities for some models by $>$50\% over baseline performance for public benchmarks and $\sim$10-16\% for our internal benchmark. This work establishes a new paradigm for evaluating interactive chat-based software engineering agents through benchmark mutation techniques. Our code is available at: \url{https://github.com/microsoft/SWE-Bench-Mutated-CAIN26}
\end{abstract}

\maketitle

\section{Introduction}

The emergence of AI-powered Software engineering agents have transformed the developer landscape. Interactive chat-based agents, including Claude Code~\cite{claude_code_2024}, VSCode Agent~\cite{vscode_2025}, represent a paradigm shift from traditional fully autonomous agents like GitHub Copilot Agent~\cite{copilot_agent_2024} that process complete problem specifications independently. Unlike their autonomous counterparts, interactive agents like Claude Code engage in iterative, contextual conversations with developers and solving the problem slowly over the course of the conversation. Therefore, evaluating them requires methodologies that capture the nuances of human-chat interactions in software development.

Current evaluation benchmarks for software engineering agents, particularly SWE-Bench Verified~\cite{swebenchverified}, suffer from some fundamental limitations when it comes to their effectiveness in assessing interactive chat-based agents. Firstly, these benchmarks are constructed from GitHub issues, which often contain detailed and well thought out problem descriptions that differ substantially from the concise, informal queries typical of IDE chat interactions. Secondly, the public availability and extensive study of these benchmarks has led to model overfitting~\cite{swebench_illusion}, where systems perform really well on public benchmarks but fail to generalize in the same way to non-public benchmarks.

To address these limitations, we introduce a benchmark mutation methodology that transforms existing formal problem descriptions into more realistic user queries based on empirical analysis of developer behavior. Our approach leverages internal telemetry data from an IDE-based agent, which we use to identify common patterns in how developers communicate bugs to chat-based assistants, then uses these patterns to systematically transform benchmark problems while preserving their essential technical content.

The contributions of this paper are as follows:

1. \textbf{Analysis of Developer Communication Patterns}: We present a novel analysis of how developers communicate bugs to chat-based agents with the goal of building such a mutation pipeline. We identify several distinct templates that characterize real-world user chat interactions.

2. \textbf{Benchmark Mutation Methodology}: We introduce a systematic approach for transforming formal benchmark problems into realistic initial user queries. As a way of testing our methodology on real-world benchmarks, we create SWE-Bench Verified-Mutated, Multi-SWE-Bench (TypeScript)-Mutated and SWE-Bench C\#-Mutated datasets. We show how our approach is highly extensible and can be applied to almost any bug-fixing benchmark.

3. \textbf{Empirical Agent Evaluation}: We provide the first comprehensive evaluation of the open-source coding agent, OpenHands~\cite{wang2024opendevinopenplatformai}, on both original and mutated benchmarks, revealing significant performance gaps and highlighting the importance of realistic evaluation scenarios.

4. \textbf{Code Artifacts}: We incorporate our approach into the official SWE-Bench GitHub repo and share our working implementation consisting of all the prompts we used and mutations generated for the SWE-Bench Verified dataset\footnote{Implementation of SWE-Bench-Mutated available at: \url{https://github.com/microsoft/SWE-Bench-Mutated-CAIN26}}. We hope that this allows the community to easily replicate our results and adopt this technique in future agent evaluations. 

Our findings demonstrate that traditional benchmarks overestimate agent capabilities by 20-50\% for publicly available datasets, while the performance gap narrows to 10-16\% for internal benchmarks like SWE-Bench C\#, showing that SWE-Bench Verified not only suffers from over-specification, but also overfitting. These results show the critical need for evaluation methodologies that accurately reflect real-world usage patterns as well as keeping benchmarks private to avoid overfitting seen in public benchmarks.

\section{Background and Related Work}
We build upon a rich foundation of research in software engineering agents, agentic benchmarks and evaluation methodology while addressing a crucial gap, which is that existing benchmarks evaluate agents using formal GitHub issues that do not reflect how developers actually communicate with chat-based coding agents in practice.
\subsection{Software Engineering Benchmarks}
SWE-bench benchmark~\cite{jimenez2024swebench} marked a turning point in software engineering agent evaluation by moving beyond isolated function / file-level tasks to real-world repo-level problems drawn from GitHub issues. This benchmark contained tasks of higher complexity than prior benchmarks, requiring agents to navigate and understand repository context, coordinate changes across multiple files, and produce patches that pass existing as well as hidden test cases. It's successor SWE-Bench Verified~\cite{swebenchverified} addressed concerns with the quality and contamination, and provided a human-verified set of 500 examples to evaluate agents against. However, problems still remain in the benchmark. Studies like SWE-Bench+~\cite{aleithan2024swebenchpluscodingbenchmark} have revealed issues like solution leakage and weak test suites. 

SWE-bench has been expanded to new domains. SWE-Bench-Multimodal~\cite{yang2024swebenchmultimodalaisystems} introduces JavaScript tasks with visual elements, while Multi-SWE-Bench~\cite{zan2025multiswebenchmultilingualbenchmarkissue} targets 7 programming languages and provides expert-annotated examples. Similarly, SWE-Poly-Bench~\cite{rashid2025swepolybenchmultilanguagebenchmarkrepository} also targets languages beyond Python. Zhang et al. introduce SWE-Bench-Live~\cite{zhang2025swebenchgoeslive}, which is a continuously updating benchmark spanning a large set of repositories that's meant to combat model overfitting and data contamination issues. While these benchmarks have advanced the area of fixing functional bugs with coding agents substantially, our work reveals a fundamental limitation they all share i.e. their reliance on formal GitHub issues fails to capture how developers actually communicate with chat-based agents in IDE environments. Our mutation methodology uniquely addresses this gap by transforming these formal benchmark problems into more realistic user queries, enabling more accurate assessment of agent performance in real-world scenarios.

\subsection{Software Engineering Agents}
Recent work has developed sophisticated frameworks for autonomous software engineering. SWE-Agent~\cite{yang2024sweagent} introduced us to the Agent-Computer Interface (ACI) design principles and achieved 12.5\% on SWE-Bench benchmark. OpenHands~\cite{wang2024opendevinopenplatformai} provides an open-source platform where agents write code, interact with the command-line, and browse the web like human developers. Interestingly, Xia et al. proposed~\cite{xia2024agentlessdemystifyingllmbasedsoftware} that uses a simple three-phase process of localization, repair, and validate and is able to solve 32\% of the problems in SWE-Bench-Lite without a complex agent architecture unlike its predecessors.

Building on reasoning capabilites of LLMs, ReAct~\cite{yao2023reactsynergizingreasoningacting} interleaves reasoning traces and task-specific actions, while Reflexion~\cite{shinn2023reflexionlanguageagentsverbal} employs reinforcement through verbal feedback by getting agent to reflect on task feedback signals. Studies like Toolformer~\cite{schick2023toolformerlanguagemodelsteach} explore the ability of LMs to teach themselves how and when to use tools through self-supervision. In a similar vein, Gorilla~\cite{patil2023gorillalargelanguagemodel} finetune smaller LMs to use tools and surpass the capabilities of existing state of the art LLMs.

While existing agent frameworks and techniques have shown impressive capabilities on public benchmarks, our work demonstrates that their reported performance may not translate to real-world settings where users communicate very differently. By evaluating on both original and mutated benchmarks, we provide the first systematic assessment of how agent performance changes when faced with realistic user queries rather than formal GitHub issues, revealing the need for agents to be designed to handle the ambiguity and informality of real user interactions.

\subsection{Developer Behaviour \& Chat-based Assistants}
Several studies have been conducted on how developers use chat-based assistants. Ziegler et al.~\cite{ziegler} measure the impact of GitHub copilot on developer productivity. Imai et al.~\cite{imai} conduct an empirical study to investigate the effectiveness of pair-programming with Copilot in comparison to human-human pair programming. Scholl et al.~\cite{scholl2024noviceprogrammersuseexperience} study how novice programmers use ChatGPT for programming exercises. McNutt et al. investigate the potential of code-assistants in Jupyter notebooks and examined design considerations for interactive environments. Li et al.~\cite{li2025unveilingrolechatgptsoftware} analyzed characteristics of ChatGPT usage by developers and found that developers tend to prefer short and task-focused interactions with the chat. Our work builds directly upon these studies of developer behavior with chat-based agents by conducting the first comprehensive analysis specifically focused on how developers communicate bugs to these systems. By connecting developer behavior studies directly to benchmark design, we establish a new paradigm for evaluation that better reflects the actual conditions under which coding agents operate.

\subsection{Data Contamination \& Mutation-Based Benchmarks}
Data contamination is a prevalent issue when it comes to public benchmarks. Approaches like LiveCodeBench~\cite{jain2024livecodebenchholisticcontaminationfree} and SWE-Bench-Live~\cite{zhang2025swebenchgoeslive} provide continuously updating benchmarks to mitigate the impacts of data contamination in static benchmarks. DyVal~\cite{zhu2024dyvaldynamicevaluationlarge} dynamically generates evaluation sets using directed acyclic graphs (DAGs). LLMEval-3~\cite{zhang2025llmeval3largescalelongitudinalstudy} provides a framework for dynamically evaluating LLMs by building on a proprietary bank of graduate-level questions that are dynamically sampled for each evaluation run.

DyCodeEval~\cite{chen2025dynamicbenchmarkingreasoningcapabilities} provides a dynamic benchmarking approach that generates diverse problem sets to ensure consistent and reliable evaluations when evaluating LLMs under contamination risks. Zhu et al.~\cite{zhu2024inferencetimedecontaminationreusingleaked} propose Inference-Time-Decontamination (ITD) to address data contamination by detecting and rewriting leaked samples without altering problem difficulties. Our benchmark mutation methodology complements these contamination-mitigation approaches while addressing a different fundamental issue i.e. the gap between how developers communicate in GitHub issues vs a chat-based agent. While previous work has focused primarily on preventing model overfitting through continuous benchmark updating or dynamic generation, our approach transforms existing benchmarks to better match real-world usage patterns. This not only helps combat overfitting, as shown by the substantial performance differences between public and private benchmarks in our results, but also enables more accurate assessment of how agents perform under realistic conditions.

\section{Methodology}

\subsection{Telemetry Analysis and Pattern Extraction}

To understand how developers actually communicate with chat-based software engineering agents, we conducted a comprehensive analysis of the internal usage telemetry data from a widely-used chat-based agent tool. This analysis formed the foundation for our benchmark mutation approach by identifying the distinct patterns that characterize real-world user interactions for fixing bugs with a chat-based agent.

\subsubsection{Data Collection and Labeling}

We collected a random sample of 10,000 user queries from a week's worth of requests issued by $\sim$6k distinct users to a chat-based agent. Since our goal is to focus on bug-fixing, we first attempt to categorize the queries into various user intents.

\begin{figure}[htbp]
\begin{lstlisting}[language={}, caption={}, label={fig:category_prompt}]
You are given a set of user queries to a software engineering agent chat within an IDE:
{sample_queries}
Please group these into 5-10 meaningful, high-level categories that capture the main themes of the kinds of software engineering tasks users issued a query for. Pick categories like "Bug Fixing", "Software Testing", "Code Search", etc. Return ONLY a numbered list of category names.
\end{lstlisting}
\caption{The prompt we use for categorizing user queries into a set of high-level categories based on the underlying task.}
\label{fig:categories_prompt}
\end{figure}

\begin{figure}[htbp]
\begin{lstlisting}[language={}, caption={}, label={fig:labeling_prompt}]
We have the following categories of tasks a user might give to a software engineering agent within an IDE chat:
1. Bug Fixing & Error Resolution
2. Code Refactoring & Quality Improvement
3. Software Testing
4. Documentation & Technical Writing
5. Feature Development & Enhancement
6. Build, Deployment and CI/CD
7. Code Navigation, Search and Analysis
8. DevOps, Infrastructure Tasks
9. Project Setup & Configuration
10. User Interface & UX Design

Assign the following user query to a software engineering agent the single best category from the list above.
Query:
{query}
Return ONLY the category name.
\end{lstlisting}
\caption{The prompt we use for labeling a user query into one of the 10 high-level categories in our data.}
\label{fig:categorize_prompt}
\end{figure}

To ensure systematic categorization, we employed a two-stage labeling process using LLMs with human validation.
\begin{enumerate}
    \item \textbf{Extract High-Level Categories in Sample}: The first stage involved generating high-level categories for software engineering tasks. We presented the LLM with a sample of $\sim$500 user queries and use the prompt shown in Figure~\ref{fig:categories_prompt} to get the LLM to generate meaningful task categories. This yielded categories in Figure ~\ref{fig:query-categories}.
    \item \textbf{Apply Categorization to New Queries}: In the second stage, we applied the categorization prompt shown in Figure~\ref{fig:categorize_prompt} to assign each query to its most appropriate category. Figure ~\ref{fig:query-categories} shows examples of such queries.
\end{enumerate}

\begin{figure}[h]
    \centering
    \includegraphics[width=\linewidth]{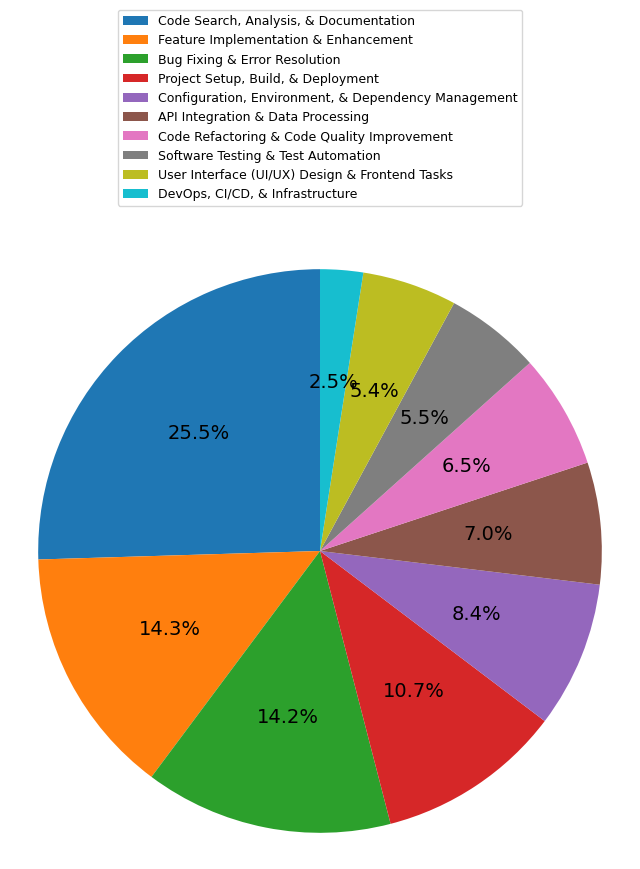}
    \caption{Distribution of High-Level categories in user queries to a coding agent. We can see that the top categories are Code Search, Analysis (Blue), Feature Implementation (Orange) and Bug Fixing (Green).}
    \label{fig:categories}
\end{figure}

Figure~\ref{fig:categories} shows the distribution of the resulting categories. We can see that bug fixing and error resolution is among the largest categories, accounting for $\sim$14\% of all queries.

\lstset{
    basicstyle=\ttfamily\footnotesize,
    breaklines=true,
    breakatwhitespace=false,
    columns=flexible,
    keepspaces=true,
    frame=none,
    xleftmargin=0pt,
    framexleftmargin=0pt,
    backgroundcolor=\color{white},
    tabsize=1,
    showtabs=false,
    showspaces=false,
    showstringspaces=false,
    basewidth=0.5em,
    gobble=0
}

\begin{figure*}[htbp]
\centering
\small
\begin{tabular}{|p{0.47\textwidth}|p{0.47\textwidth}|}
\hline
\textbf{A. API Integration \& Data Processing} & \textbf{B. Bug Fixing \& Error Resolution} \\
\hline
\parbox[t]{0.45\textwidth}{\ttfamily\footnotesize\obeylines%
I have an agent called IncidentClassificationAgent and I want to expose this agent's functionalities via MCP}
& 
\parbox[t]{0.45\textwidth}{\ttfamily\footnotesize\obeylines%
When I run the following command:

./ib\_player --rank 2 --numberOfRanks 4 --ipAddresses <addresses> --listenPorts 2000 2256 2512 2768 --schedule /mnt/store/user/player/file.bin --replayCount 2 --gidIndex 3 --device dev0

\vspace{8pt}
I get the following error:

[PLAYER][ALWAYS] GOAL IB Player

[PLAYER][ALWAYS] Number of IP addresses: 4

terminate called after throwing an instance of `std::runtime\_error'

  what():  Binary file contains invalid magic cookie: \%d4402

Aborted (core dumped)

\vspace{8pt}
Please help me find where the error is occuring.}
\\
\hline
\textbf{C. Code Refactoring \& Code Quality} & \textbf{D. Code Search, Analysis, \& Documentation} \\
\hline
\parbox[t]{0.45\textwidth}{\ttfamily\footnotesize\obeylines%
can you not use a HashMap, and instead use a struct with Average and Max, to avoid having the dynamic memory allocation on each pod}
& 
\parbox[t]{0.45\textwidth}{\ttfamily\footnotesize\obeylines%
tell me about the YELLOW GDM or PDM, what is it, what do we need it for, what is the business impact and what exactly we need to build?}
\\
\hline
\textbf{E. Configuration \& Dependency Management} & \textbf{F. DevOps, CI/CD, \& Infrastructure} \\
\hline
\parbox[t]{0.45\textwidth}{\ttfamily\footnotesize\obeylines%
can you install docker in wsl?}
& 
\parbox[t]{0.45\textwidth}{\ttfamily\footnotesize\obeylines%
debug this pipeline run: https://dev.azure.com/user/...

get logs using cli}
\\
\hline
\textbf{G. Feature Implementation \& Enhancement} & \textbf{H. Project Setup, Build, \& Deployment} \\
\hline
\parbox[t]{0.45\textwidth}{\ttfamily\footnotesize\obeylines%
I want to code a Real-time defect detection using AI vision systems to ensure high-quality output. code me a python project powered by GPT 4 multimodal
\vspace{2pt}
}
& 
\parbox[t]{0.45\textwidth}{\ttfamily\footnotesize\obeylines%
I need to setup an AI experimentation platform for a mid sized team for a POC and eventual production rollout, help me choose what to deploy}
\\
\hline
\textbf{I. Software Testing \& Test Automation} & \textbf{J. User Interface (UI/UX) Design \& Frontend} \\
\hline
\parbox[t]{0.45\textwidth}{\ttfamily\footnotesize\obeylines%
Review my changes to client.go, inspect and understand client\_test.go and add / modify existing tests to account for the changes made}
& 
\parbox[t]{0.45\textwidth}{\ttfamily\footnotesize\obeylines%
I want to create a tick tac toe game that can run in browser. 2 players should be able to play it on same computer. Make the game look beautiful with modern design}
\\
\hline
\end{tabular}
\caption{Categorization of user queries to a software engineering agent into 10 high-level categories. We show an example for each category of user query.}
\label{fig:query-categories}
\end{figure*}

\lstset{
    basicstyle=\ttfamily\scriptsize,
    breaklines=true,
    frame=single,
    backgroundcolor=\color{gray!10},
    keywordstyle=\color{blue},
    commentstyle=\color{green!60!black},
    stringstyle=\color{red},
    showstringspaces=false,
    captionpos=b
}

\begin{figure}[h]
    \centering
    \includegraphics[width=\linewidth]{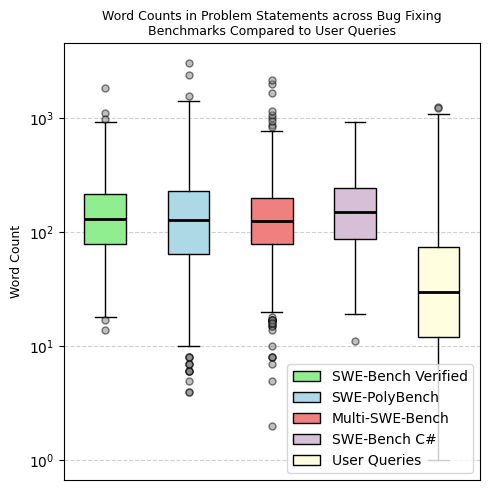}
    \caption{Distribution of word counts in the problem statements of different benchmark compared to real-world user queries. The distributions show how much more concise telemetry queries tend to be compared to bug-fixing benchmarks.}
    \label{fig:word_count}
\end{figure}

\begin{figure}[h]
    \centering
    \includegraphics[width=\linewidth]{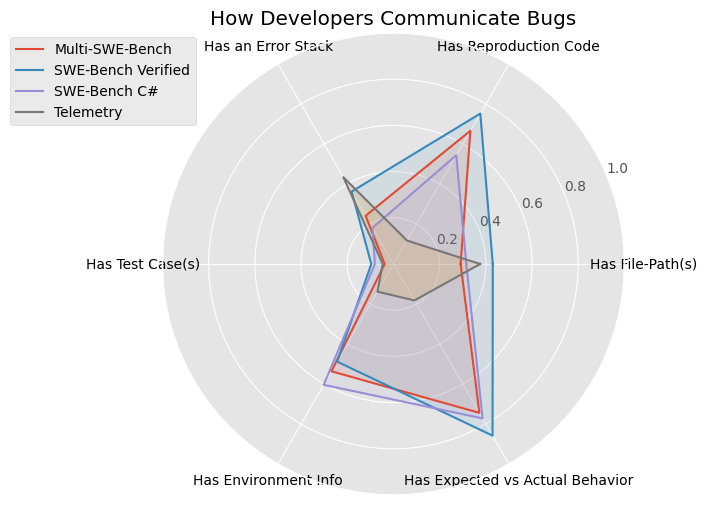}
    \caption{Plot showing a comparison of how developers communicate in benchmarks vs real-world user queries to chat-based agents. We can see that telemetry queries contain very different kinds of information compared to GitHub issues.}
    \label{fig:how_devs_communicate}
\end{figure}

\subsubsection{Understanding Bug-Fix User Queries}
\label{templates}
To better understand the characteristics of the bug-fixing subset of user queries, we compare them against problems in various bug-fixing benchmarks for coding agents. Figure \ref{fig:word_count} shows a side by side comparison of word counts in the problem statements of public bug-fixing benchmarks such as SWE-Bench Verified~\cite{swebenchverified}, SWE-PolyBench~\cite{rashid2025swepolybenchmultilanguagebenchmarkrepository} and Multi-SWE-bench~\cite{zan2025multiswebenchmultilingualbenchmarkissue} against bug-fixing user queries in telemetry. We see a dramatic difference in query length distributions. We can see that user queries tend to be lot more succinct than benchmarks, being in the range of 10-30 words, while problems from benchmarks like SWE-Bench typically contain $>$100 words.

\begin{figure}[htbp]
\begin{lstlisting}[language={}, caption={}, label={}]
You are analyzing GitHub issue reports for software bug datasets.

Given the text of a GitHub issue, your task is to determine whether it contains the following types of information. 
Answer strictly in JSON format.

For each field, answer "yes" or "no" based only on the content of the issue.

Definitions:
- has_file_path: The issue explicitly mentions a source file path or file name relevant to the bug (e.g. "in models/ridge.py" or "src/utils/helpers.cpp").
- has_reproduction_code: The issue includes code snippets, commands, or minimal reproduction examples that demonstrate how to reproduce the bug.
- has_error_stack: The issue contains an error message, traceback, or stack trace output (e.g. "ValueError: invalid shape").
- has_test_case: The issue includes a test (unit test or assertion) that fails or demonstrates the bug.
- has_environment_info: The issue specifies the runtime environment, OS, version numbers, dependencies, or configuration details.
- has_expected_vs_actual: The issue clearly distinguishes expected behavior versus actual behavior (e.g. "expected X but got Y").

Return your answer exactly in the following JSON format:

{{
  "has_file_path": "yes" | "no",
  "has_reproduction_code": "yes" | "no",
  "has_error_stack": "yes" | "no",
  "has_test_case": "yes" | "no",
  "has_environment_info": "yes" | "no",
  "has_expected_vs_actual": "yes" | "no"
}}

Now analyze the following GitHub issue:

---
{issue_text}
---
\end{lstlisting}
\caption{The prompt we use to determine whether a problem statement contains key elements that make a bug report useful, such as file-path(s), reproduction code, test case(s), expected vs actual behavior and environment / configuration details.}
\label{fig:elements_prompt}
\end{figure}

To further demonstrate the differences between the communication styles of users, we measure how often benchmarks and user queries from telemetry contain various key elements that make a bug report useful, such as file-path(s), reproduction code, test case(s), expected vs actual behavior and environment / configuration details, using prompt shown in Figure \ref{fig:elements_prompt} with an LLM. Figure \ref{fig:how_devs_communicate} shows a comparison of how often these elements show up in the queries contained within benchmarks, containing primarily GitHub issues, and user queries. We can see that users tend to communicate more with the error stack and file-paths compared to other elements in agent queries. In stark contrast, GitHub issues contain reproduction code, expected behavior, environment details, etc. as well. These features are largely missing from user queries. Our analysis reflects chat users' preference for sharing more targeted information like file-paths, error stacks, rather than providing exhaustive details upfront as typically seen in GitHub issues.

\begin{figure}[htbp]
\begin{lstlisting}[language={}, caption={}, label={}]
You are analyzing how users communicate software bugs to an AI coding assistant. I will give you a list of real user queries where the user is trying to fix a bug. Your task is to carefully read through these examples and identify the common templates or patterns they follow.

By "templates or patterns" I mean recurring ways in which the bug is described or reported. Examples include:
- Pasting the call stack or error message without explanation
- Asking the agent to fix a specific function or line of code
- Describing the expected vs. actual behavior
- Asking for debugging or diagnosis without providing code

Instructions:
- Read all the provided queries.
- Group them into common templates/patterns.
- For each template:
  - Give it a short, descriptive name.
  - Provide 5 example queries from the dataset that match it.

Here are the user queries:
{queries}
\end{lstlisting}
\caption{The prompt we use to assign extract templates / patterns from bug-fixing user queries.}
\label{fig:template_prompt}
\end{figure}

We also conduct a deeper analysis to identify specific communication patterns within these bug-fixing queries to understand how users communicate bugs to a chat-based agent. To systematically identify communication patterns, we employed LLMs with the template extraction prompt shown in Figure~\ref{fig:template_prompt}.

This yielded 11 distinct communication templates that capture how developers report bugs to chat-based agents:

\begin{enumerate}
    \item \textbf{Paste Error/Stack Trace Only}: In these cases, the user simply pastes an error message, call stack, crash report, or test failure log with little to no added context or explanation.
    \item \textbf{Direct Fix This/Fix the Error}: User asks for a fix without detailed explanation, often using concise phrases such as “fix this,” “fix the error,” or “resolve the error”, etc.
    \item \textbf{Paste Error + Ask for Explanation}: User pastes an error or stack trace and requests help understanding, diagnosing, or analyzing the root cause.
    \item \textbf{Minimal Fix/Change Request}: User briefly shows an error, command, or warning and directly asks for the correction, expected syntax, or minimal code change.
    \item \textbf{Specific Line/Function Fix}: In this case, user requests help to fix a particular function, line number, file, or a single narrow area in code.
    \item \textbf{Expected vs. Actual Behavior}: User clearly describes what should happen (expected) and what is happening (actual), asking why there is a mismatch or how to resolve the difference. 
    \item \textbf{Paste Code with Question}: User pastes a code snippet and then includes a clarifying question, request for investigation, or asks what is wrong or how to fix.
    \item \textbf{Request for Debugging Without Code}: User wants diagnosis, investigation, or suggestions for a bug described only in words or with minimal technical data/logs (no code or error pasted).
    \item \textbf{Test/CI Failure Paste}: User pastes test output (including success/failure logs), often from CI/CD runs, to ask why a test fails or how to fix.
    \item \textbf{Request for Analysis/Triage}: User requests a triage, diagnosis, or deep-dive analysis of a bug or incident, sometimes referencing IDs or tickets, often without pasting technical details.
    \item \textbf{Referencing Documentation/External Issue}: User refers to an external issue tracker, document, or specification and requests analysis or a fix.
\end{enumerate}

Each template represents a distinct cognitive approach to problem communication, reflecting different levels of technical detail, context provision, and expected agent response.

\begin{figure}[h]
    \centering
    \includegraphics[width=\linewidth]{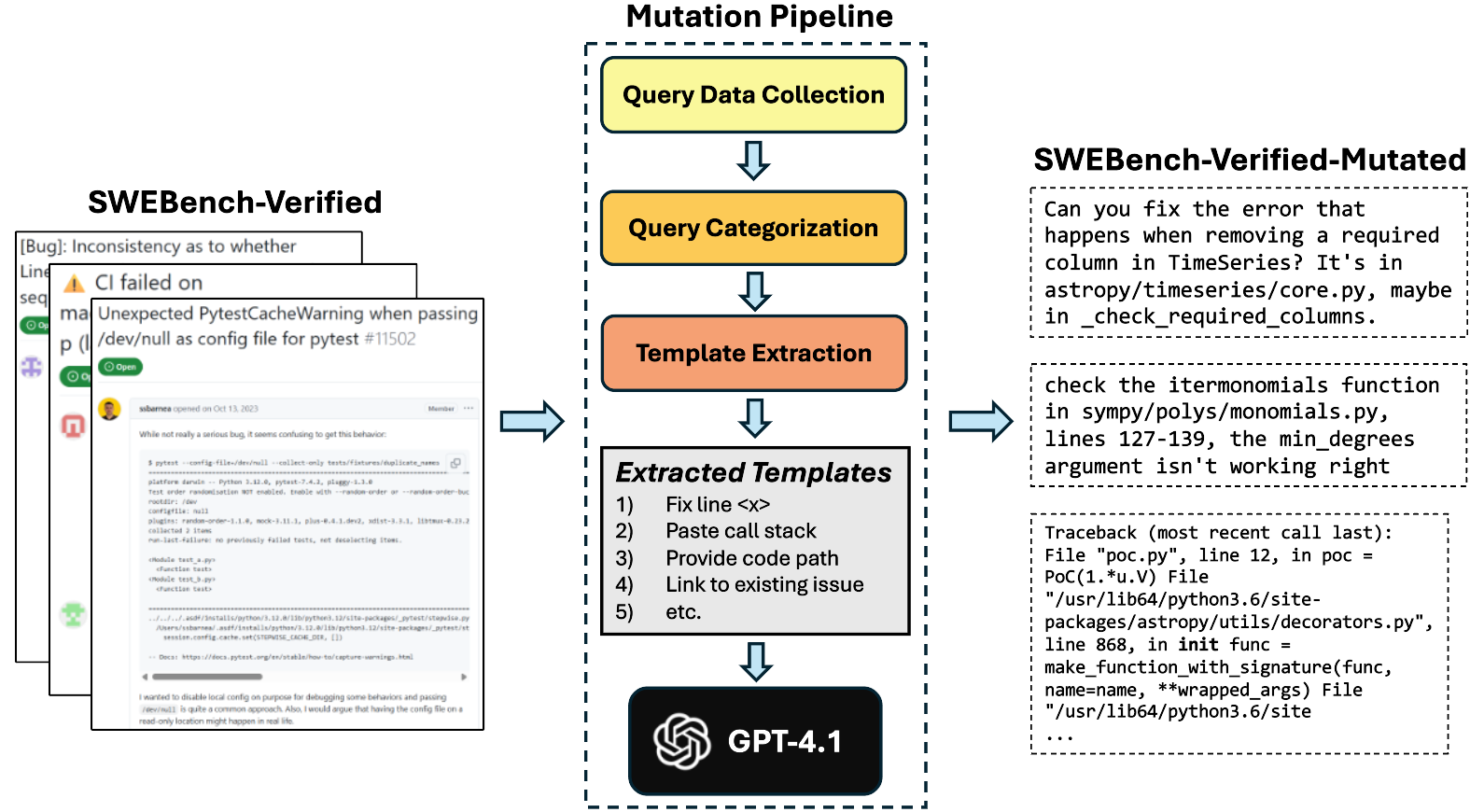}
    \caption{A high-level overview of our benchmark mutation approach.}
    \label{fig:mutation_pipeline}
\end{figure}

\subsection{Mutation Methodology}

Building on our template analysis, we developed a systematic methodology for transforming formal benchmark problems into queries matching real-world usage patterns. Figure~\ref{fig:mutation_pipeline} shows the high-level pipeline of our approach. We describe our mutation algorithm in more detail below and then give an example of a mutation generated for SWE-Bench.

\begin{figure}[htbp]
\centering
\begin{minipage}[t]{0.48\textwidth}
\textbf{\small Original Issue:}\hfill\texttt{\footnotesize astropy\_\_astropy-7336}
\begin{mdframed}[style=issuebox]
\scriptsize
\textbf{Summary}
I am using the \texttt{units.quantity\_input} decorator with typing hints for constructors, however when I add the correct return value for the constructor (\texttt{None}) then I get an exception, because \texttt{None} has no attribute \texttt{to}.

\vspace{0.5em}
\textbf{Reproducer}

The issue can be reproduced with the following file:

\begin{lstlisting}[language=Python]
import astropy.units as u
class PoC(object):
    @u.quantity_input
    def __init__(self, voltage: u.V) -> None:
        pass
if __name__ == '__main__':
    poc = PoC(1.*u.V)
\end{lstlisting}

which results in the following error:

\begin{lstlisting}
$ python3 poc.py
Traceback (most recent call last):
  File "poc.py", line 12, in <module>
    poc = PoC(1.*u.V)
  File "/usr/lib64/python3.6/site-packages/astropy/utils/decorators.py", line 868, in __init__
    func = make_function_with_signature(func, name=name, **wrapped_args)
  File "/usr/lib64/python3.6/site-packages/astropy/units/decorators.py", line 225, in wrapper
    return return_.to(wrapped_signature.return_annotation)
AttributeError: 'NoneType' object has no attribute 'to'
\end{lstlisting}

This has been tested on Fedora 27 with python 3.6.3, astropy 2.0.2 and numpy 1.13.3 all from Fedora's repository.

\vspace{0.5em}
\textbf{Workaround}

The issue can be circumvented by not adding the return type typing hint. Unfortunately, then a static type checker cannot infer that this function returns nothing.

\vspace{0.5em}
\textbf{Possible fix}

Maybe the decorator could explicitly check whether None is returned and then omit the unit check.
\end{mdframed}
\end{minipage}%
\hfill
\begin{minipage}[t]{0.46\textwidth}
\vspace{1em}
\textbf{\small Mutated Issue:}\hfill\texttt{\footnotesize astropy\_\_astropy-7336-mutated}
\scriptsize
\begin{lstlisting}
Traceback (most recent call last):
    File "poc.py", line 12, in <module>
      poc = PoC(1.*u.V)
    File "/usr/lib64/python3.6/site-packages/astropy/utils/decorators.py", line 868, in __init__
     func = make_function_with_signature(func, name=name, **wrapped_args)
    File "/usr/lib64/python3.6/site-packages/astropy/units/decorators.py", line 225, in wrapper return return_.to(wrapped_signature.return_annotation) 
AttributeError: 'NoneType' object has no attribute 'to'
\end{lstlisting}
\end{minipage}
\caption{We show a comparison of a mutated example generated by our approach for issue \texttt{astropy\_\_astropy-7336} in SWE-Bench Verified. The box on the top shows the body of the original issue, which is akin to a form the user must fill out when submitting an issue to the astropy repo. It contains elements like reproduction code, call-stack along with a description of the issue, workaround, etc. The box at the bottom contains the mutated issue output by our approach, which only contains the call-stack part of the original issue. This matches the pattern `Paste Error / Stack Trace Only' in Section \ref{templates}}
\label{fig:mutation-issue-comparison}
\end{figure}

\subsubsection{Mutation Algorithm}

\begin{figure*}[htbp]
\begin{lstlisting}[language={}, caption={}, label={}]
You are given:
- A set of transformation templates showing how people informally describe bugs to an interactive coding assistant.
- A software bug description from the SWE-Bench Verified dataset (in GitHub issue style).
- The corresponding code patch that fixes the bug.

Your task:
- Apply as many transformation templates as make sense for this example.
- For each applicable template, rewrite the bug description in the style of that template.
- Use the patch details (e.g., file paths, function names, and line numbers) where relevant to make the request realistic.
- Make the query realistic to how users may query a chat-based agent. Users tend to write short, incomplete descriptions, often with typos. This is very IMPORTANT!!
- Skip templates that clearly do not apply.
- Output each transformed description as a separate bullet point in the following format:
  **Transform Kind 1**
  TEXT
  **Transform Kind 2**
  TEXT
  ------

Transformation templates:
{templates}
Bug description:
"""{problem}"""
Code patch:
"""{patch}"""
Now generate all applicable transformed descriptions.
\end{lstlisting}
\caption{The prompt we use to mutate problems in a bug-fixing benchmark to match the style of communication seen in user queries. We provide the LLM with the following 3 inputs: the problem statement, the code patch and the templates generated by our templatization algorithm.}
\label{fig:mutation_prompt}
\end{figure*}

We use LLMs for applying the mutation with the prompt shown in Figure~\ref{fig:mutation_prompt}. The prompt takes as inputs: a problem description from an existing benchmark, the corresponding code patch that fixes the problem, and the set of communication templates extracted from telemetry analysis in Section~\ref{templates}. The algorithm asks the model to apply multiple applicable templates to generate a diverse and realistic set of variants for each problem. The reason why our mutation prompt incorporates patch information is to enable realistic references to specific files, functions, and line numbers that users often include in their queries.

For each benchmark problem, the mutation process generates multiple potential user queries representing different communication approaches. We randomly select one variant per problem to create the final mutated benchmark, ensuring diversity while maintaining consistency in evaluation.

\subsubsection{Mutation Example From SWE-Bench}
Figure \ref{fig:mutation-issue-comparison} shows an example of a potential mutation generated by our approach for an issue in SWE-Bench Verified. This mutation is for instance \texttt{astropy\_\_astropy-7336} from astropy repo. We can see that the orignal issue contains a form filled out by the user when they submitted the issue, with fields like issue description, reproduction code, error stack, workaround, etc. Our model recognizes that this isn't how a user would communicate with a chat-based agent and one of the mutations it generates is shown in the bottom box. The mutation is to simply extract the call stack from the issue body itself and use that as the problem statement. This follows the pattern ``Paste Error/Stack Trace Only" seen in our template analysis. We refer to the resulting instance as \texttt{astropy\_\_astropy-7336-mutated}.

\begin{figure*}[htbp]
    \centering
    \begin{subfigure}[b]{0.32\textwidth}
        \centering
        \includegraphics[width=\textwidth]{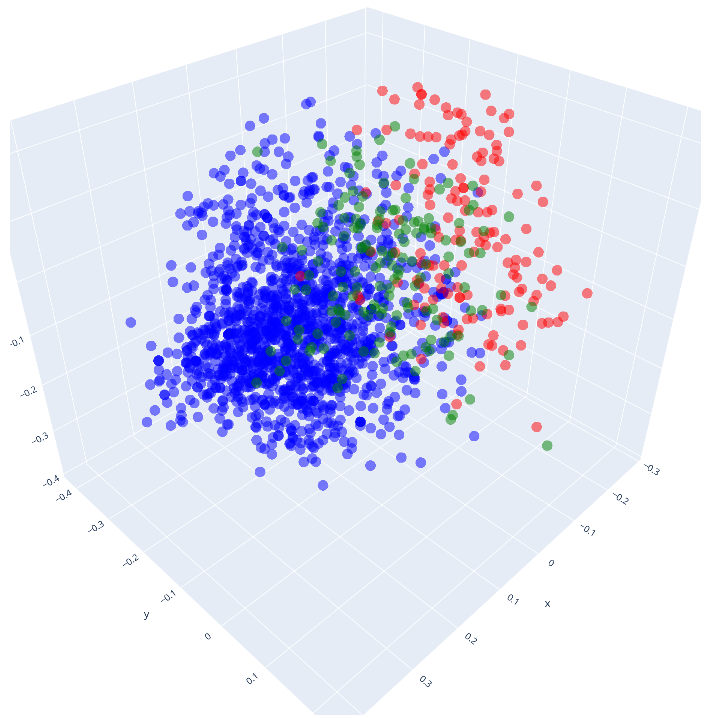}
        \label{fig:view1}
    \end{subfigure}
    \hfill
    \begin{subfigure}[b]{0.32\textwidth}
        \centering
        \includegraphics[width=\textwidth]{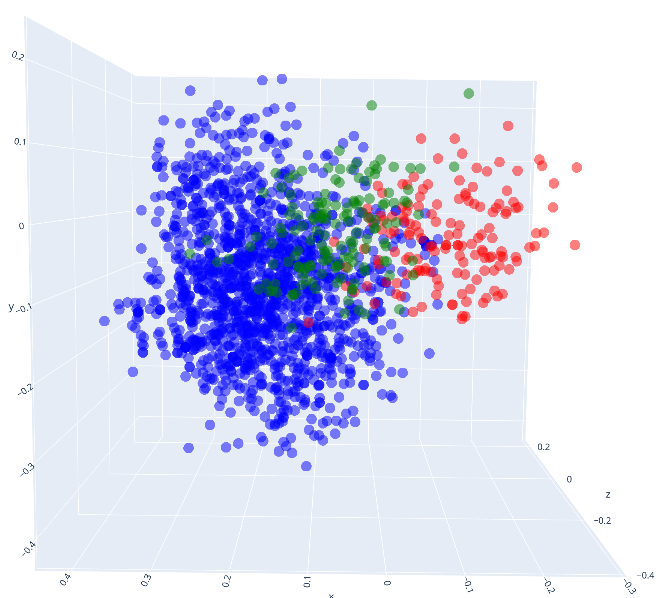}
        \label{fig:view2}
    \end{subfigure}
    \hfill
    \begin{subfigure}[b]{0.32\textwidth}
        \centering
        \includegraphics[width=\textwidth]{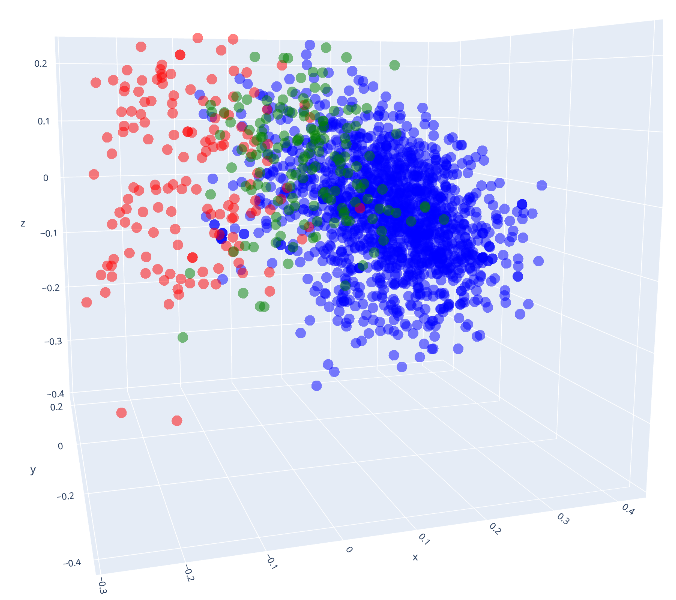}
        \label{fig:view3}
    \end{subfigure}

    \begin{tikzpicture}
        \node[anchor=south east] at (8.5,0) { 
            \begin{tikzpicture}
                \draw[thick, rounded corners] (0,0) rectangle (3.8,1.45);
                \node[fill=blue!70, circle, inner sep=2pt] at (0.3,1.1) {};
                \node[right] at (0.6,1.15) {\footnotesize Telemetry queries};
                
                \node[fill=red!70, circle, inner sep=2pt] at (0.3,0.7) {};
                \node[right] at (0.6,0.75) {\footnotesize SWE-Bench C\#};
                
                \node[fill=green!40!black, circle, inner sep=2pt] at (0.3,0.3) {};
                \node[right] at (0.6,0.35) {\footnotesize SWE-Bench C\#-Mutated};
            \end{tikzpicture}
        };
    \end{tikzpicture}
    
    \caption{Views of point clouds corresponding to queries from different sources, after being embedded using OpenAI's \texttt{text-embedding-3-large} model. The three views show different rotations of the same 3D plot containing telemetry queries (blue), SWE-Bench C\# (red), and SWE-Bench C\#-Mutated (green) queries. We can see that the queries corresponding to the mutated benchmark overlap more with the cloud corresponding to telemetry data.}
    \label{fig:3d-point-cloud}
\end{figure*}

\subsubsection{Visualizing Mutated Queries Against Telemetry}
Figure \ref{fig:3d-point-cloud} shows a 3d visualization where we plot the mutations generated by our approach for SWE-Bench C\# dataset, as well as the original benchmark, against our set of telemetry queries. All the queries were embedded using \texttt{text-embedding-3-large} model, followed by PCA to reduce the dimensionality of the embeddings to be plotted in a 3d figure. We can see that the embeddings corresponding to SWE-Bench C\# (red) and the telemetry queries (blue) make two separate clouds, with little to no overlap. However, the cloud corresponding to SWE-Bench C\#-Mutated (green) has a lot more overlap with telemetry query cloud than the original benchmark. This shows that our mutation approach successfully transforms benchmark queries to better match the characteristics of actual developer interactions. While the original benchmark queries form a distinct cluster separated from real-world usage patterns, our mutation strategy shifts the distribution toward more realistic user queries. This visualization provides further evidence that our approach produces queries that more closely resemble how developers naturally communicate with chat-based agents in practice, validating the core premise of our benchmark mutation methodology.

\section{Experimental Setup}
In this section, we describe our experimental setup to assess the impact of mutating the benchmark with the above methodology.

\subsection{Benchmarks}
We mutate the following three bug-fixing benchmarks:
\begin{itemize}
    \item \textbf{SWE-Bench Verified}: This benchmark consists of 500 human-verified bug-fixing problems drawn from popular Python repositories. This is commonly used for evaluating coding agents today and, therefore, a natural choice. Each problem within this benchmark includes a problem statement based on a GitHub issue, repository context, and ground-truth solution patch. We mutate each problem in this benchmark into SWE-Bench Verified-Mutated.
    \item \textbf{SWE-Bench C\#}: We also mutate a bug-fixing benchmark internal to our organization. It consists of 150 bug-fixing problems drawn from C\# repositories on GitHub in a similar fashion to SWE-Bench. These problems are of varying level of difficulties and consist of the GitHub issue, repo context and the developer patch. We call the mutated version of this benchmark SWE-Bench C\#-Mutated.
    \item \textbf{Multi-SWE-Bench (TypeScript)}: We mutate the TypeScript subset of Multi-SWE-Bench. We chose to focus only on a subset (224 examples) due to the size of the overall dataset. We call the mutated version of this benchmark Multi-SWE-Bench (TypeScript)-Mutated.
\end{itemize}

For each benchmarks, we apply the mutation prompt in Figure \ref{fig:mutation_prompt} and generate multiple mutations for each problem statement. For each example, we randomly pick one mutation from the set of possible mutations generated by the model.

\subsection{Agent \& Model Configurations}
We use the open-source agent OpenHands~\cite{wang2024opendevinopenplatformai} for our evaluation. We use OpenHands' default prompting for bug fixing tasks, with the only exception being the change in wording from Python to C\#, for our C\# benchmarks. We evaluate the agent against the following three commonly used LLMs models: GPT-4.1, Claude Sonnet 3.7 and Claude Sonnet 4.

Each agent run is allowed a maximum of 100 steps to complete an example in a given benchmark, after which agent success is assessed through the same harness used to evaluate the baseline unmutated benchmark. All examples are run in isolated docker containers.

\subsection{Metrics}
For each task above, we report the following metrics:
\begin{itemize}
    \item \textbf{Success Rate (\%)}: Percentage of tasks where the agent produces a fix that passes the hidden tests within the harness for a given benchmark, baseline or mutated.
    \item \textbf{Avg. Token Usage (\# of tokens)}: In addition to correctness metrics, we also report the number of input and output tokens used by agent per task.
    \item \textbf{Avg. Steps (\# of steps)}: Even though we give the agent a maximum of 100 steps, it's possible for it to terminate before it exhausts the max steps. We report the average steps taken by the agent to complete a task.
\end{itemize}

While a higher success rate indicates better chances of task completion, it often comes with increased agent runtime and computational/monetary cost (higher token usage and step count). The user often wants to balance quality and cost, so we analyzed the effects to both classes of metrics.

\begin{table*}[t]
\centering
\caption{Comparison of success rate, steps and tokens used by OpenHands Agent on mutated and baseline variants of SWE-Bench Verified, SWE-Bench C\# and Multi-SWE-Bench (TypeScript).}
\label{tab:consolidated_results}
\begin{subtable}[t]{0.31\textwidth}
\centering
\fontsize{6}{7}\selectfont
\caption{SWE-Bench Verified}
\begin{tabular}{l|c|c|c}
\toprule
\textbf{Metric} & \textbf{GPT-4.1} & \textbf{Sonnet 3.7} & \textbf{Sonnet 4} \\
\midrule
\multicolumn{4}{c}{\textbf{Success Rate (\%)}} \\
\midrule
Baseline & 35.6 & 54.2 & 65.4 \\
Mutated & 22.6 & 39.2 & 50.2 \\
Change (\%) & \impact{-36.5}{red} & \impact{-27.8}{red} & \impact{-23.2}{red} \\
\midrule
\multicolumn{4}{c}{\textbf{Avg. Steps}} \\
\midrule
Baseline & 39.4 & 38.0 & 44.6 \\
Mutated & 44.0 & 43.2 & 47.8 \\
Change (\%) & \impact{+11.6}{red} & \impact{+13.6}{red} & \impact{+7.2}{red} \\
\midrule
\multicolumn{4}{c}{\textbf{Avg. Tokens}} \\
\midrule
Baseline & 825k & 1.36M & 1.38M \\
Mutated & 721k & 1.14M & 1.25M \\
Change (\%) & \impact{-12.6}{green} & \impact{-16.2}{green} & \impact{-9.4}{green} \\
\bottomrule
\end{tabular}
\end{subtable}%
\hspace{0.02\textwidth}
\begin{subtable}[t]{0.31\textwidth}
\centering
\fontsize{6}{7}\selectfont
\caption{SWE-Bench C\#}
\begin{tabular}{l|c|c|c}
\toprule
\textbf{Metric} & \textbf{GPT-4.1} & \textbf{Sonnet 3.7} & \textbf{Sonnet 4} \\
\midrule
\multicolumn{4}{c}{\textbf{Success Rate (\%)}} \\
\midrule
Baseline & 16.7 & 27.3 & 32.7 \\
Mutated & 14.0 & 22.7 & 29.2 \\
Change (\%) & \impact{-16.2}{red} & \impact{-16.8}{red} & \impact{-10.7}{red} \\
\midrule
\multicolumn{4}{c}{\textbf{Avg. Steps}} \\
\midrule
Baseline & 48.4 & 51.8 & 54.3 \\
Mutated & 47.9 & 54.1 & 57.3 \\
Change (\%) & \impact{-1.0}{green} & \impact{+4.4}{red} & \impact{+5.5}{red} \\
\midrule
\multicolumn{4}{c}{\textbf{Avg. Tokens}} \\
\midrule
Baseline & 961k & 1.85M & 2.13M \\
Mutated & 940k & 1.96M & 2.18M \\
Change (\%) & \impact{-2.2}{green} & \impact{+5.9}{red} & \impact{+2.3}{red} \\
\bottomrule
\end{tabular}
\end{subtable}
\hspace{0.02\textwidth}
\begin{subtable}[t]{0.31\textwidth}
\centering
\fontsize{6}{7}\selectfont
\caption{Multi-SWE-Bench (TypeScript)}
\begin{tabular}{l|c|c|c}
\toprule
\textbf{Metric} & \textbf{GPT-4.1} & \textbf{Sonnet 3.7} & \textbf{Sonnet 4} \\
\midrule
\multicolumn{4}{c}{\textbf{Success Rate (\%)}} \\
\midrule
Baseline & 16.1 & 25.4 & 15.6 \\
Mutated & 13.4 & 18.8 & 7.2 \\
Change (\%) & \impact{-16.8}{red} & \impact{-26.0}{red} & \impact{-53.8}{red} \\
\midrule
\multicolumn{4}{c}{\textbf{Avg. Steps}} \\
\midrule
Baseline & 65.7 & 42.3 & 52.2 \\
Mutated & 67.2 & 43.5 & 58.8 \\
Change (\%) & \impact{+2.2}{red} & \impact{+2.8}{red} & \impact{+12.6}{red} \\
\midrule
\multicolumn{4}{c}{\textbf{Avg. Tokens}} \\
\midrule
Baseline & 1.35M & 1.32M & 1.87M \\
Mutated & 1.40M & 1.37M & 2.14M \\
Change (\%) & \impact{+3.7}{red} & \impact{+3.8}{red} & \impact{+14.4}{red} \\
\bottomrule
\end{tabular}
\end{subtable}
\end{table*}

\section{Results}
Our experimental evaluation reveals insights into the real-world performance of chat-based software engineering agents and the impact of realistic problem presentation on benchmark outcomes.

\subsection{Impact on Success Rate}
\subsubsection{SWE-Bench Verified Results}
Table~\ref{tab:consolidated_results} shows the results of running OpenHands agent against baseline and mutated SWE-Bench Verified. We see that the mutation of SWE-Bench Verified problems from formal GitHub issues to realistic user queries results in a substantial performance degradation across all agent-model combinations. For all models, the agent suffers a 20-40\% drop in relative success rates compared to baseline when it encounters realistic user queries instead of a longer GitHub issue style problem description.

These results demonstrate that existing benchmarks significantly overestimate agent's bug fixing capabilities. The consistent performance drops across models suggest that the gap is not simply due to model-specific limitations, but rather reflects fundamental differences in problem presentation and agent adaptation to realistic scenarios. We see a higher drop for GPT-4.1 compared to other models. This could be because this model perhaps requires more context, such as in a GitHub issue, to solve a given problem or is possibly over-fitted.
\subsubsection{SWE-Bench C\# Results}
In contrast to the Python benchmark results, mutation of our internal C\# benchmark yields a much smaller performance impact, as seen in Table~\ref{tab:consolidated_results}. The smaller absolute performance drops (2-5\% versus 13-16\% in SWE-Bench Verified) in agent-model performances could be for a number of reasons. One possible reason could be that LLMs or the agent are overfitting to popular benchmarks like SWE-Bench Verified. Another could be that the agent's performance on C\# isn't very high to begin with. This could be because LLMs are simply more adept at writing python than C\#. This would make sense as Python repos are more abundant than C\#~\cite{gitcharts}, so there is more training data available for these models. Not just model, agents could also be overfitted to Python development, since they were built with Python benchmarks like SWE-Bench in mind. 

\subsubsection{Multi-SWE-Bench (TypeScript) Results}
Table~\ref{tab:consolidated_results} (c) shows the results for running the agent on the TypeScript subset of Multi-SWE-Bench with and without mutations. Once again we see drops in agent performance for all 3 models when the bug descriptions are mutated. In fact, the relative performance degradation for TypeScript benchmark exceeds the degradation observed in the C\# benchmark in all 3 models.
Interestingly, we see the largest drop in the results for Sonnet 4. The model also performs worse compared to other models with the baseline benchmark as well.

\subsection{Impact on Steps \& Token Usage}
\subsubsection{SWE-Bench Verified Results}
Looking at Table~\ref{tab:consolidated_results}, we see a rise in the number of steps taken by the agent. This is expected because we've taken away a lot of the details in the GitHub issue, so the agent has to deliberate for longer or search for relevant code before arriving to a solution. However, there is a drop in the token usage of the agent when we mutate SWE-Bench Verified. This is interesting because we would expect the agent to take more tokens proportional to the rise in the number of steps. We believe that this is because the agent has fewer details to work with --- each step in the trajectory becomes less verbose.


One more interesting insight into the numbers for all benchmarks is that Claude Sonnet 4 seems to be more verbose and takes more steps than the other two models, showing the model's tendency to engage with the problem longer compared to the other models.

\subsubsection{SWE-Bench C\# Results}
We see that in this case, the number of steps doesn't rise quite as drastically as the SWE-Bench Verified results. This could again be an artifact of overfitting. Claude Sonnet 4 seems to suffer the highest increase in the number of steps, followed by Sonnet 3.7. Interestingly, the number of steps taken by GPT-4.1 actually goes down, albeit slightly. We should note that the performance of the agent with this model is also lower, so the drop in steps may not entirely be a positive sign.

Unlike SWE-Bench Verified, we see a rise in token usages from the Claude models and roughly equal token usage from the GPT model compared to the baseline. This makes sense because the models are having to reason more about the under-specified problem, which takes more steps and tokens.

If we compare the token usages of the models between SWE-Bench Verified and SWE-Bench C\# benchmarks, we see that Claude Sonnet 4 seems to use almost twice as many tokens per instance for the C\# benchmark compared to the python benchmark. Neither of the other models have a similar rise in token usage. This might be another case of language parity in LLMs, where they perform disproportionally better for one language over the other or are simply overfitting to a benchmark.
\subsubsection{Multi-SWE-Bench (TypeScript) Results}
We see the number of steps taken by the agent to solve a given task rises for all the models when mutations are applied. Similar to the success rate, the rise in steps and token usage is also the most drastic for Sonnet 4. Looking more closely at the agent trajectories, we notice that for both mutated and unmutated benchmarks the agent exhausts max steps in $\sim$3 times as many instances for Sonnet 4, compared to Sonnet 3.7. This shows Sonnet 4 tends to struggle with TypeScript bug fixing in general.

Similar to SWE-Bench C\#, Sonnet 4 also consumes a lot more tokens and steps for TypeScript bug-fixing tasks, unlike tasks in SWE-Bench Verified. GPT-4.1 also takes more steps and tokens compared to other benchmarks.



\section{Limitations}

While our study provides valuable insights into mutating benchmarks for realistic agent eval, we acknowledge several limitations and present opportunities for future research.

\textbf{Language / Task Coverage}: Firstly, our work focuses primarily on benchmarks for Python, C\# and TypeScript languages. Expanding to other programming languages will help strengthen the generalizability of our approach. Additionally, the benchmarks we examine only cover bug-fixing tasks. As demonstrated in our analysis of user queries to chat-based agents, bug fixing accounts for less than 15\% of all queries. Our methodology's efficacy for other coding tasks (feature implementation, code refactoring, software testing, etc.) remains unexplored. In future work, we plan to investigate generalizing this approach to other categories of tasks that developers commonly use coding agents for.

\textbf{Agent Coverage}: We evaluate only one open-source agent (OpenHands) on our mutated benchmarks. While this represents a reasonable choice given that OpenHands is among the state-of-the-art research agents~\cite{swebench-leaderboard}, other agents may have different capabilities and exhibit other performance patterns when used with mutated benchmarks. Furthermore, our evaluation framework focuses on single-interaction benchmarks and does not account for the iterative, conversational nature of real chat-based agents, where users can provide clarifications and additional context through follow-up exchanges.

\textbf{Template Extraction \& Reliance on LLM-based Analysis}: Although we believe the templates we identified are comprehensive for our dataset and usage patterns, developers may communicate differently with other coding agents with different design and UI elements, whose telemetry we don't have access to. Domain-specific communication patterns may not be represented in our sample. Additionally, our LLM-based categorization, template extraction, and application approach, despite human validation, may introduce systematic biases due to our choice of model and prompt design.

\textbf{Problem Validity Following Mutation}: While our mutation process aims to mimic realistic user behavior when interacting with coding agents, it may inadvertently discard essential technical information in some of these benchmark problem statements and alter problem difficulty. Our approach relies on the assumption that problems stay valid even after the mutation. However, for the modified problems, multiple solutions besides the original developer fix may be valid, potentially leading to unfairly penalizing the agents for reasonable solutions to an under-specified problems.

\textbf{Unchanged Evaluation Methodology \& Harness}: Though we modify problem statements to reflect realistic user queries, the eval harness we use remains unchanged and rigid. Ideally, we would develop a better set of metrics that capture the nuances of human-agent interaction and better approximate user satisfaction with agent responses.

Despite these limitations, our work establishes a foundational approach toward realistic eval for bug-fixing capabilities in chat-based coding agents and provides an extensible framework that can be adapted to existing and new benchmarks.

\section{Conclusion}

In this study, we have demonstrated that existing bug-fixing benchmarks for evaluating software engineering agents, such as SWE-Bench Verified, Multi-SWE-Bench, systematically overestimate agent capabilities by $>$50\% over baselines, due to a combination of their heavy reliance on formal GitHub issue descriptions, language parity and overfitting. By analyzing real-world developer interactions with chat-based coding agents and extracting templates of how user communicate bugs to a chat, we introduced a benchmark mutation methodology that transforms problems from GitHub issue style communication to a more realistic chat-style query. Our empirical results and analysis show that this approach reveals significant performance gaps in agents and agent evaluation methodologies, highlighting the importance of realistic evaluation scenarios and the risks of benchmark overfitting. We advocate for the adoption of a mutation-based evaluation for agents and private benchmarks to better reflect actual developer usage and provide a realistic measure of agent capabilities while avoiding the pitfalls of overfitting. We believe future work should explore extending this methodology to other software engineering tasks as well as domains and further refine this technique to capture the nuances of multi-turn user-agent collaboration.

\bibliographystyle{ACM-Reference-Format}
\bibliography{iclr2024_conference}


\end{document}